\documentclass[11pt]{article}
\usepackage{amsfonts,amscd,amsmath,amssymb,graphicx}

\def\th{T_{\text{\tiny H}}}
\def\rt{r_{\text{\tiny T}}}
\def\g{g_{\text{\tiny eff}}}
\def\v{V_{\text{\tiny X}}}

\def\ep{\text{e}}
\def\pl{{\cal P}}

\title{String Free Energy, Hagedorn and Gauge/String Duality}
\author{Oleg Andreev\thanks{Also at Landau Institute for Theoretical Physics, Moscow. }\\\\
{\it Technische Universit\"at M\"unchen, Excellence Cluster,} \\
{\it Boltzmannstrasse 2, 85748 Garching, Germany}}
\date{}
\begin{document}
\vspace{-8cm}
\maketitle
\begin{abstract}
We give examples of modeling the string free energy whose behavior mimics that of QCD: a power-law at high temperature and an exponential 
decrease at low temperature. Although the effective description is in terms of strings, no limiting temperature exists, as expected for a crossover.
 \end{abstract}

\vspace{-10.5 cm}
\begin{flushright}

{\small SPAG-A2/08 }\\
\end{flushright}

\vspace{9.5cm}


\section{The Problem}
Since the late sixties and early seventies, a limiting temperature also known as the Hagedorn temperature
$\th$ \cite{hagedorn} has attracted great interest in string theory \cite{list}. At zero temperature, strings show an exponential growth in 
the density of states that is a good point for strings to be applied for describing the physics of hadrons. However, the growth is so rapid that the 
partition function of a string gas converges only for temperatures below $\th$, which indicates the breakdown of the string picture 
at higher $T$. There were many attempts to treat $\th$ as a temperature associated with a phase transition. 
The idea is that at low $T$ the QCD spectrum may be described by strings and above $\th$ by a gas of quarks and gluons. 

However, the recent results from lattice QCD at zero baryonic chemical potential $\mu_{B}$ indicate that
in the real world there is no phase transition but an analytic crossover \cite{cross}.\footnote{It is believed that this is the case for nonzero 
values of $\mu_{B}$ up to a few hundred MeV. For a review, see \cite{stephanov}.} This changes the story completely. If strings are indeed 
relevant for QCD then one has to show that a stringy description is also valid for high $T$. 

In this paper, we propose two possible models for getting the high temperature behavior of strings which looks like that in QCD. Good 
motivations for this are models which recently helped to resolve another somewhat similar puzzle: How does string theory reproduce the hard 
behavior of gauge theory scattering amplitudes? In particular, in \cite{ps} an elegant solution 
based on Maldacena duality \cite{ads} was suggested. The essence of this solution is the warped geometry of the string dual. Later, in 
\cite{asiegel}, string models with quantized tension were proposed to obtain string amplitudes with Regge poles and parton behavior. 
So we are bound to learn something if we succeed.

\section{Possible Resolutions}

\subsection{Flux Tube Picture}

There is a long history, going back to Nambu's conjecture, of modeling QCD-like states using flux tubes. In particular, extensive numerical 
simulations have demonstrated that the flux picture does occur: a string-like chromoelectric flux tube forms between distant static 
color charges, as shown in Figure 1 at left.\footnote{See, e.g., \cite{lattice} for a review.} On the other hand, it turns out that QCD at short distances 
is in violent disagreement with expectations from a fluctuating string. The common wisdom is that, in order for the flux picture 
\begin{figure}[ht]
\begin{center}
\includegraphics[width=8cm]{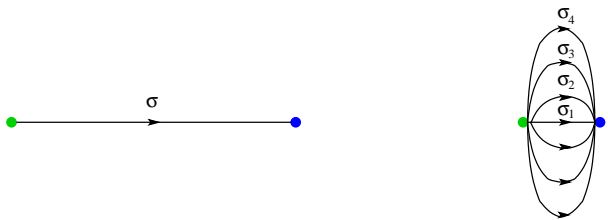}
\caption{\small{Left: a typical string-like flux tube of tension $\sigma$. Right: a fat string as a collection of thin strings with different tensions.}}
\end{center}
\end{figure}
to be of use, the flux must be much longer than its intrinsic width. Since at short distances a string, as observed in lattice QCD, gets fat, this 
completely destroys the string picture. 

Before we look at more detailed modeling in QCD, let us begin by thinking about the flux tubes in QED. The Faraday's concept of lines of flux 
emanating from charged bodies is a nice way to visualize electric and magnetic fields. For two opposite charges, it looks like that 
shown in Figure 1 at right. If we assume that such a picture holds in QCD at short distances, then we come to a conclusion that fluxes of different 
length should have different tensions. Otherwise, we end up with a single flux shown in Figure 1 at left which has the minimal energy. The 
next step is to interpret the fat string of lattice QCD as a set of thin strings of different tensions $\{\sigma_n\}$, as illustrated in Figure 1 at right. 

Having established a starting point for the flux picture at short distances, we turn our attention to more detailed properties of it. First,  in the 
case of continuous spectrum of string tensions it seems natural to promote $\sigma$ to a new spacetime coordinate. If so, then strings effectively 
live in some warped 5-dimensional space with $\sigma$ playing a role of the fifth dimension. In fact, it was Polyakov who first realized that 
in order to describe 4-dimensional gauge theories dual string theories must be 5-dimensional \cite{polyakov}. Note that in doing so, he 
interpreted the Liouville field as the fifth coordinate and introduced the notion of running tension. In contrast, we used a more phenomenological 
way based on the Coulomb-type heavy quark potential at short distances.\footnote{We are grateful to V.I. Zakharov for numerous discussions 
of these issues.} Second, another possibility is a discrete spectrum of string tensions. This line of thinking was pursued 
in \cite{asiegel}.\footnote{An earlier proposal of a dual resonance model, but with quantized Regge slope, is due to \cite{goto}.}
Certainly, this is not the whole story. Spectra of more realistic models might contain both discrete and continues pieces. 

\subsection{Quantized Tension}

We begin with an ensemble of strings whose tensions take discrete values (quantized) \cite{asiegel}. This implies that the original string 
parameter $\alpha'$ is promoted to a discrete function $\alpha'_n$ with $n$ a positive integer. The explicit form of $\alpha'_n$ is restricted 
by requiring that the mass spectrum of the $n$-th string is a subset of that of the primary string. The simplest possible proposal for the free energy 
of such an ensemble seems to be that $F$ is a sum of string free energies\footnote{This is similar in spirit to what was proposed for scattering 
amplitudes in \cite{asiegel}. Note that it assumes no interaction between strings with different tensions.}

\begin{equation}\label{FE}
F=\sum_{n=1}^\infty w_n F(n)
\,.
\end{equation}
Here the first term $F(1)$ is associated with a primary string such that $\alpha'_1=\alpha'$ and $w_1=1$. Then, the contribution of the 
$n$-string $F(n)$ is obtained from $F(1)$ by replacing $\alpha'\rightarrow\alpha'_n$.

To simplify things slightly and illustrate the essential point, we will consider $F(1)$ as a partition function of the free string gas and choose 
$\alpha'_n=\alpha'/\pl_k(n)$ with $\pl_k(n)$ being a polynomial of degree $k$ such that $\pl_k(n)$ takes only integer values and grows 
with $n$. With this choice, we have for higher states\footnote{Here, for simplicity, we consider the bosonic case.}

\begin{equation}\label{Z}
\sum_n w_n
\int^\infty [dh]\,\ep^{4\pi h}\ep^{-h\sqrt{\pl_k}/\sqrt{\alpha'}T}
\,.
\end{equation}
The first term of the series diverges for temperatures greater than the Hagedorn temperature $\th=1/4\pi\sqrt{\alpha'}$. The divergence 
occurs because the density of states (entropy factor) grows exponentially. This is the well known story. The novelty is that now one can 
get rid of it by imposing a new lower bound for the spectrum of the string tensions.  Indeed, if 

\begin{equation}\label{bound-n}
n_\ast <n
\,,
\end{equation}
with $n_\ast$ a solution to the equation
\begin{equation}\label{nast}
\pl_k(n)=T^2/\th^2
\,,
\end{equation}
then one finds that the exponent in \eqref{Z} is negative and the integral over $h$ is convergent. Notice that there is no effect on the lower bound 
if $T<\th$ because $\pl_k(1)=1$. Also, \eqref{bound-n} has a simple physical meaning: it suppresses long strings in the ensemble which are 
responsible for the divergence and, as a result, no limiting temperature exists.

This discussion suggests that we have to refine the proposal for the free energy. In the present case, we can take

\begin{equation}\label{FE-r}
F=\sum_{\max\{1, 1+[n_\ast]\}}^\infty w_n F(n)
\,,
\end{equation}
where $[n_\ast]$ means the integer part of $n_\ast$.

Now we will investigate the behavior of $F$ at high temperature. In this case, eq.\eqref{nast} simplifies to $a_k n^k=T^2/\th^2$ which has a 
solution $n_\ast=\sqrt[k]{T^2/a_k\th^2}$. We can replace the sum by an integral and $[n_\ast]$ by $n_\ast$ that is also applicable for large $n$. 
Assuming that $w_n$ is a product of power functions like $n^\delta \pl_k^\gamma$, eq.\eqref{FE-r} then becomes

\begin{equation}\label{FEhT}
F=\frac{1}{\sqrt{\alpha'}}\int^\infty_{n_\ast} dn\,n^\delta \pl_k^\gamma\,F\bigl(\sqrt{\alpha'/\pl_k}\,T\bigr)
\,.
\end{equation}
 We have included the factor $1/\sqrt{\alpha'}$ on dimensional grounds. To see that the integral provides the desired power-law behavior, we 
rescale $n$ as $n\rightarrow \sqrt[k]{T^2/\th^2}\,n$. As a result, we have\footnote{Note that $F_0$ must be finite, otherwise \eqref{FEhT2} 
makes no sense.}

\begin{equation}\label{FEhT2}
F=F_0 T^{2\gamma+2(\delta+1)/k}\Bigl[1+O\Bigl(\frac{1}{T}\Bigr)\Bigl]
\,.
\end{equation}
The parameters can be fixed by fitting the high temperature behavior in QCD. We take the option $\delta=-1$ and $\gamma=2$. It is 
worth noting that these  values also lead to the correct partonic behavior of scattering amplitudes \cite{asiegel} \footnote{Note that the shift 
$\gamma\rightarrow\gamma +2$ is due to scattering of vector particles in \cite{asiegel}.} that provides a cross check of the model.

The last issue concerning the string model with the quantized tension that we will discuss here is the low
temperature behavior. If the primary string has no tachyon and massless modes in its spectrum (or, in other words, it has a mass gap), then for 
low $T$ the free energy of the $n$-th string is given by $F(n)\sim T^a\ep^{-m_n/T}$ with $m_n\sim \pl_k(n)/\sqrt{\alpha'}$. Since $\pl_k$ 
is an increasing function of $n$, the sum \eqref{FE-r} is dominated by the term with $n=1$, and as a result, the free energy exponentially 
decreases at low temperature\footnote{We will not need the detail expression for $A$ in this discussion.}

\begin{equation}\label{lowT}
F=A\ep^{-\frac{m}{T}}
\,,
\end{equation}
where $m=m_1$.

Thus, in the two different limits we have modeled precisely the temperature behavior as that of QCD at zero baryonic chemical potential.

\subsection{Warped Geometry}

Since a string dual of QCD is unknown, for illustrative purposes, we consider a string theory whose space-time is a product of the 
Schwarzschild black hole in $\text{AdS}_5$ with a compact five-dimensional space $\text{X}$

\begin{equation}\label{metric}
ds^2=\frac{r^2}{R^2}\bigl(l dt^2+d\vec{x}^2\bigr)+\frac{R^2}{r^2}l^{-1}dr^2+ds_{\text X}^2
\,,\quad l=1-\frac{\rt ^4}{r^4}\,,
\quad
\rt=\pi R^2 T
\,,
\end{equation}
where $t$ is periodic with period $\beta=1/T$. It is believed that such a black hole geometry does capture important properties of QCD at high 
$T$ \cite{witten}. Because of the warping, in local inertial coordinates the period, for a state approximately localized in the $r$-direction, is 
given by

\begin{equation}\label{period}
\tilde\beta=\sqrt{l}\frac{r}{R}\,\beta
\,,
\quad
\text{with}
\quad
\rt <r<\infty
\,.
\end{equation}
The essential point is that a single ten-dimensional temperature $\tilde T=1/\tilde\beta $ can give rise to different values of a four-dimensional 
temperature $T$. Indeed, for given $\tilde\beta$, \eqref{period} becomes

\begin{equation}\label{Tr}
T=\alpha r
\,,
\end{equation}
with $\alpha^2=-\tfrac{1}{2}\tilde\beta^2+\sqrt{\pi^4R^4+\tfrac{1}{4}\tilde\beta^4}/\pi^4R^6$. We see that
$T$ grows with $r$ and approaches infinity for $r\rightarrow\infty$. This allows us to study the string free energy
at high $T$ even for low ten-dimensional temperature such that $\tilde T<\th$. The latter, for example, means that the partition function of 
the free string gas remains finite and no limiting temperature occurs. Thus, the warping might be a point in describing the crossover of 
QCD in terms of strings.\footnote{Aside from hard scattering \cite{ps}, where a single ten-dimensional momentum $\tilde p$ also gives rise to 
different values of  a four-dimensional momentum $p$ via $p=\frac{\tilde p}{R} r$, the warping is also the essence of the 
Randall-Sundrum proposal for solving the hierarchy problem \cite{rs}.}

 To write an expression for the string free energy on backgrounds like \eqref{metric} in the large-$r$ region,
  we treat it as a ten-dimensional expression at fixed $r$, integrated coherently over this position\footnote{Such a treatment seems natural for the 
states localized in the $r$-direction. In the context of scattering amplitudes, it was proposed in \cite{ps}.}

\begin{equation}\label{fenergy}
\frac{F}{T}=-\ln Z=\frac{1}{\alpha'{}^5}\sum_{i=1}^\infty \g^{2(i-1)} \int d^{10}x \sqrt{g} \,F^{(i)} (\sqrt{\alpha'}\tilde T)
\,,
\end{equation}
where $\g$ is an effective coupling and  $F^{(i)}$ is a $i$-loop contribution \cite{aw}. We have included the factor $\frac{1}{\alpha'{}^5}$ on 
dimensional grounds. Using \eqref{metric} and \eqref{period}, we find the free energy density

\begin{equation}\label{highT}
f=\frac{F}{V}=\frac{1}{\alpha'{}^5}\frac{\v}{ R^3}
\sum_{i=1}^\infty \g^{2(i-1)} \int^\infty_{\pi R^2 T} dr\,r^3 \,F^{(i)} \bigl(\sqrt{\alpha'} RT/\sqrt{l} r\bigr)
\,,
\end{equation}
 where $V$ is a $3d$ spatial volume and $\v$ is a volume of the internal space. The integration limits are due to \eqref{period}. From this 
expression it is evident that a simple rescaling $r\rightarrow  T r$ leads to 

\begin{equation}\label{fenergy2}
f\sim T^4
\,.
\end{equation}
The scaling of this free energy density with $T$ is precisely as in QCD.

Now let us look at the lower limits of the integrals in \eqref{FEhT} and \eqref{highT}. In \eqref{FEhT},
$n_\ast$ comes from the bound \eqref{bound-n} which suppresses long strings and yields the convergence of the partition function of the free 
string gas; the lower limit in \eqref{highT} is due to the horizon of the black hole \eqref{metric}. Does it mean that the horizon suppresses long 
strings? We can gain some understanding of this by choosing $\pl_2(n)=n^2$ and discretizing the $r$-direction as $r=nR$. Then, 
\eqref{bound-n} can be rewritten as $T/\th <n$. On the other hand, $\rt <r$ now becomes $c(\lambda ) T/4\th < n$, 
where $c=R/\sqrt{\alpha'}$ is a function 
of the 't Hooft coupling $\lambda$. We see that the horizon resolves the Hagedorn singularity if the coupling is large enough, such 
that $c(\lambda )>4$.\footnote{Formally, we may consider only the term $F^{(1)}$ in \eqref{highT}.}

We conclude this section by making a few remarks. First, our derivation assumes that the right hand side of \eqref{highT} is finite. Even if we 
avoid the problem of the Hagedorn singularity in ten dimensions, there is no guarantee for convergence. Unfortunately, we cannot, with our 
present methods, perform accurate calculations in type IIB string theory on $\text{AdS}_5$  and, therefore, address the issue of convergence. 
Second, it is straightforward to extend the above analysis to any background whose metric is given by \eqref{metric} at large $r$ and, as a result,
 recover the desired scaling of the free energy at high $T$. Unfortunately, there is no satisfactory modeling for a string background which is 
dual to  QCD at low $T$. So, we leave the analysis of this limit for future study. Finally, as in the case of high-energy string scattering in 
warped spacetime \cite{a}, the power law behavior at high temperature can be understood from the analysis of bosonic zero modes in a 
string path integral. These modes contribute the volume factor $\int d^{10}\sqrt{g}$ which does lead to the scaling of the free energy 
with temperature like in QCD.

\section{Concluding Comments}

Although at first glance, both approaches look similar, they are really different. In the first approach the starting point is an 
ensemble of strings which has already a good behavior at low $T$. There is no limiting (Hagedorn) temperature because long strings are 
suppressed at high $T$. In the second approach one gets high $T$ in a gauge theory from a ten-dimensional string dual whose temperature 
$\tilde T$ is below $\th$. The effect is due to the warping, where a single ten-dimensional temperature gives rise to many different values of
a four-dimensional temperature and, as a result, there is no need to go beyond $\th$ on the string theory side to get high $T$ on the gauge theory 
side.

It is necessary to stress that our analysis goes beyond the supergravity approximation. It is quite general and may be also applied to 
other supersymmetric and heterotic string theories.

For large $N_c$, a gauge theory free energy is of order $N_c^2$. If the sum in \eqref{fenergy} is an $1/N_c$ expansion, then it has no 
contribution from zero genus as it starts from genus one. This seems puzzling. A possible resolution is that zero genus somehow effectively 
occurs after resummation. To see that this is indeed the case, we need the full control of type IIB string theory on curved backgrounds like 
$\text{AdS}_5$ that is unfortunately beyond our grasp at present. 

As in \cite{ps, asiegel}, we have been able to recover the partonic behavior at high energy (temperature) without revealing the nature 
of partons as well as the logarithmic scaling violation. A more complete description remains an interesting avenue for future research.

\vspace{.15cm}
{\bf Acknowledgments}

\vspace{.15cm}
\noindent We are grateful to J. Kapusta, M. Strassler and S. Sugimoto for discussions, and W. Siegel for
encouragement. We also would like to thank S. Hofmann and A.A. Tseytlin for comments and reading the manuscript. This work 
was supported in part by DFG "Excellence Cluster'' and the Alexander von Humboldt Foundation under Grant No. PHYS0167. 
We acknowledge the warm hospitality at the Institute for Nuclear Theory at the University of Washington, where a main portion of this 
work was done.
 


\small
\begin{thebibliography}{99}
\bibitem{hagedorn}
R. Hagedorn, Nuovo Cimento Suppl. {\bf 3}, 147 (1965) and {\bf 6}, 311 (1968). Note that a limiting temperature has also been
considered by Yu.B. Rumer, Zh.Eksp.Teor.Fiz. {\bf 8}, 1899 (1960).
\bibitem{list}
The following is an incomplete list: S. Fubini and G. Veneziano, Nuovo Cimento {\bf 64A}, 811 (1969);
C. Lovelace, in Proceedings of the Conference on Regge Poles, Irvine, California, (1969);
K. Huang and S. Weinberg, Phys.Rev.Lett. {\bf 25}, 895 (1970);
S.C. Frautschi, Phys.Rev.D {\bf 3}, 2821 (1971);
D. Gross, J. Harvey, E. Martinec, and R. Rohm, Nucl.Phys.B {\bf 256}, 253 (1985);
J.J. Atick and E. Witten, Nucl.Phys.B {\bf 310}, 291 (1988);
B. Sundborg, Nucl.Phys.B {\bf 573}, 349 (2000).
\bibitem{cross}
Y. Aoki, G. Endrodi, Z. Fodor, S.D. Katz and K.K. Szabo, Nature {\bf 443}, 675 (2006).
\bibitem{stephanov}
M.A. Stephanov, PoS(LAT2006) 024.
\bibitem{ps}
J. Polchinski and M.J. Strassler, Phys.Rev.Lett. {\bf 88}, 031601 (2002).
\bibitem{ads}
J. Maldacena, Adv.Theor.Phys. {\bf 2}, 231 (1998); S.S. Gubser, I.R. Klebanov, and A.M. Polyakov, Phys.Lett.B {\bf 428}, 105 (1998); E. Witten, 
Adv.Theor. Math.Phys. {\bf 2}, 253 (1998).
\bibitem{asiegel}
O. Andreev and W. Siegel, Phys.Rev.D {\bf 71}, 086001 (2005).
\bibitem{lattice}
 K.J. Juge, J. Kuti, and C. Morningstar,  From surface roughening to QCD string theory, in Proceedings of the 24th Johns Hopkins 
Workshop "Nonperturbative QFT Methods and Their Applications'', Budapest, Hungary, 2000, p.143.
\bibitem{polyakov}
A.M. Polyakov, Nucl.Phys.Proc.Suppl. {\bf 68}, 1 (1998).
\bibitem{goto}
S. Naka and M. Kenmoku, Prog.Theor.Phys. {\bf 55}, 844 (1976).
\bibitem{witten}
E. Witten, Adv.Theor.Math.Phys. {\bf 2}, 505 (1998).
\bibitem{rs}
L. Randall and R. Sundrum, Phys.Rev.Lett. {\bf 83}, 3370 (1999); {\bf 83}, 4690 (1999).
\bibitem{aw}
J.J. Atick and E. Witten, as cited in \cite{list}.
\bibitem{a}
O. Andreev, Phys.Rev.D {\bf 70}, 027901 (2004).
\end{thebibliography}
\end{document}